\documentclass{article}
\usepackage{amsmath}
\usepackage{epsfig}
\usepackage{amssymb}
\usepackage{fullpage}
\usepackage{color}

\def\##1{{\underline #1}}
\def\=#1{\underline{\underline{#1}}}

\def\+#1{\underline{\bf #1}}
\def\*#1{\underline{\underline{\bf #1}}}

\def\eps{\epsilon}
\def\epso{\epsilon_{\scriptscriptstyle 0}}
\def\muo{\mu_{\scriptscriptstyle 0}}

\def\.{\mbox{ \tiny{$^\bullet$} }}

\def\le{\left(}
\def\ri{\right)}

\def\lec{\left\{}
\def\ric{\right\}}

\def\l#1{\label{#1}}
\def\r#1{(\ref{#1})}
\def\c#1{\cite{#1}}

\def\eps{\epsilon}
\def\epso{\epsilon_0}
\def\muo{\mu_0}

\def\.{\mbox{ \tiny{$^\bullet$} }}

\def\le{\left(}
\def\ri{\right)}

\def\lec{\left\{}
\def\ric{\right\}}

\def\l#1{\label{#1}}
\def\r#1{(\ref{#1})}

\begin{document}

\begin{center}

{\bf {\LARGE Towards an experimental realization of affinely transformed  linearized QED 
 vacuum via inverse homogenization}}

\vspace{10mm} \large

 Tom G. Mackay\footnote{E--mail: T.Mackay@ed.ac.uk}\\
{\em School of Mathematics and
   Maxwell Institute for Mathematical Sciences\\
University of Edinburgh, Edinburgh EH9 3JZ, UK}\\
and\\
 {\em NanoMM~---~Nanoengineered Metamaterials Group\\ Department of Engineering Science and Mechanics\\
Pennsylvania State University, University Park, PA 16802--6812,
USA}\\
 \vspace{3mm}
 Akhlesh  Lakhtakia\footnote{E--mail: akhlesh@psu.edu}\\
 {\em NanoMM~---~Nanoengineered Metamaterials Group\\ Department of Engineering Science and Mechanics\\
Pennsylvania State University, University Park, PA 16802, USA}\\
and\\
 {\em Materials Research Institute\\
Pennsylvania State University, University Park, PA 16802, USA}

\end{center}

\vspace{15mm}

\begin{abstract}

Within the framework of quantum electrodynamics (QED), vacuum is a
nonlinear medium which can be linearized for a rapidly time-varying
electromagnetic field with a small amplitude subjected to a
magnetostatic field. The linearized  QED vacuum is a uniaxial
dielectric-magnetic medium for which the degree of anisotropy is
exceedingly small. By implementing an affine transformation of the
spatial coordinates, the degree of anisotropy may become
sufficiently large as to be readily perceivable. The inverse
Bruggeman formalism can be implemented to specify a homogenized
composite material (HCM) which is electromagnetically equivalent to
the affinely transformed QED vacuum. This HCM can arise from
remarkably simple component materials;  for example, two isotropic
dielectric  materials and two isotropic magnetic materials, randomly
distributed as oriented spheroidal particles.

\end{abstract}

\vskip 1 cm

\noindent \textbf{Keywords:} quantum electrodynamics, vacuum
birefringence, homogenization, inverse Bruggeman formalism

\vskip 1 cm

\section{Introduction}

Classical vacuum  is a linear  medium for which the principle of
superposition holds. Consequently, light propagation in classical
vacuum is unaffected by the presence of a magnetostatic field.
However, within the framework of quantum electrodynamics (QED),
vacuum is a nonlinear medium \c{Jackson}. The QED vacuum can be
linearized for a rapidly time-varying electromagnetic field with a
small amplitude subjected to a slowly varying (or static) magnetic
field \c{Adler2007}. A consequence of linearization is that the QED
vacuum appears as a uniaxial dielectric--magnetic medium for optical
fields \c{Adler71}. The constitutive parameters which characterize
this uniaxial medium depend on the magnitude and direction of the
magnetostatic field.

The degree of anisotropy associated with the QED vacuum is
exceedingly small. Consequently,  a direct measurement of this
attribute poses enormous challenges to experimentalists \c{Iac}, and
experimental verification of the anisotropy of the QED vacuum is
eagerly awaited  \c{Zav}. In  view of this difficulty, we propose an
experimental simulation of the QED vacuum which would enable the
anisotropy  to be explored for practicable  magnetostatic fields.
The simulation is based on a homogenized composite material (HCM),
which arises from the homogenization of remarkably simple component
materials. For example, the component materials could be isotropic
dielectric and magnetic materials, randomly distributed as oriented
spheroidal particles. Similar HCM-based simulations have recently
been described for  the Schwarzschild-(anti-)de Sitter spacetime
\c{ML_PRB} and   cosmic strings \c{ML_cloak}.

In the following sections, 3-vectors are underlined with the
addition of the $\hat{}$ symbol denoting a unit vector. Double
underlining indicates a 3$\times$3 dyadic with the dyadic transpose
being labelled with the additional
 ${}^T$ symbol. The 3$\times$3 identity dyadic represented as $\=I$.
 The permittivity and permeability of
 classical vacuum are written as  $\epso=8.854\times10^{-12}$~F~m$^{-1}$
and $\muo=4\pi\times 10^{-7}$~H~m$^{-1}$,  respectively.

\section{Electrodynamics of QED vacuum}

 We consider vacuum under the influence of a magnetostatic field $\#B_s=
|\#B_s| \,\hat{\#B}_s$. In classical vacuum, the passage of light
 is unaffected by $\#B_s$, as reported by an inertial observer.
However, this is not the case for the QED vacuum. The QED vacuum is
a nonlinear  medium which can be linearized for rapidly
time--varying plane waves. Thereby, for propagation of light,  QED
vacuum is represented by the anisotropic dielectric--magnetic
constitutive relations \c{Adler71}
\begin{equation}
\left. \begin{array}{l} \#D = \epso \=\eps \. \#E \vspace{4pt} \\
\#B = \muo \=\mu \. \#H \end{array} \right\}. \l{CRs}
\end{equation}
 The relative permittivity and
 relative permeability dyadics of the QED vacuum have the uniaxial forms \c{LM07}
\begin{equation}
\left. \begin{array}{l} \=\eps = \le 1-8 a |\#B_s|^2 \ri  \le \=I-
\hat{\#B}_s \hat{\#B}_s \ri + \le 1+20 a |\#B_s|^2 \ri \hat{\#B}_s
\hat{\#B}_s \vspace{8pt} \\ \=\mu = \displaystyle{\frac{1}{ 1-8 a
|\#B_s|^2 } \le \=I- \hat{\#B}_s \hat{\#B}_s \ri + \frac{1}{ 1-24 a
|\#B_s|^2 } \hat{\#B}_s \hat{\#B}_s}
\end{array} \right\}, \l{CDs}
\end{equation}
where the constant $a = 6.623 \times 10^{-26}\,
{\mbox H}^{-1}~ {\mbox {kg}}^{-1}\,\,{\mbox{m}}^2\,\,{\mbox{s}}^2$.
The constitutive dyadics \r{CDs} were derived by Adler \c{Adler71}
from the Heisenberg--Euler effective Lagrangrian of the
electromagnetic field \c{HE,Schwinger}.

\begin{figure}[!ht]
\centering
\includegraphics[width=4.3in]{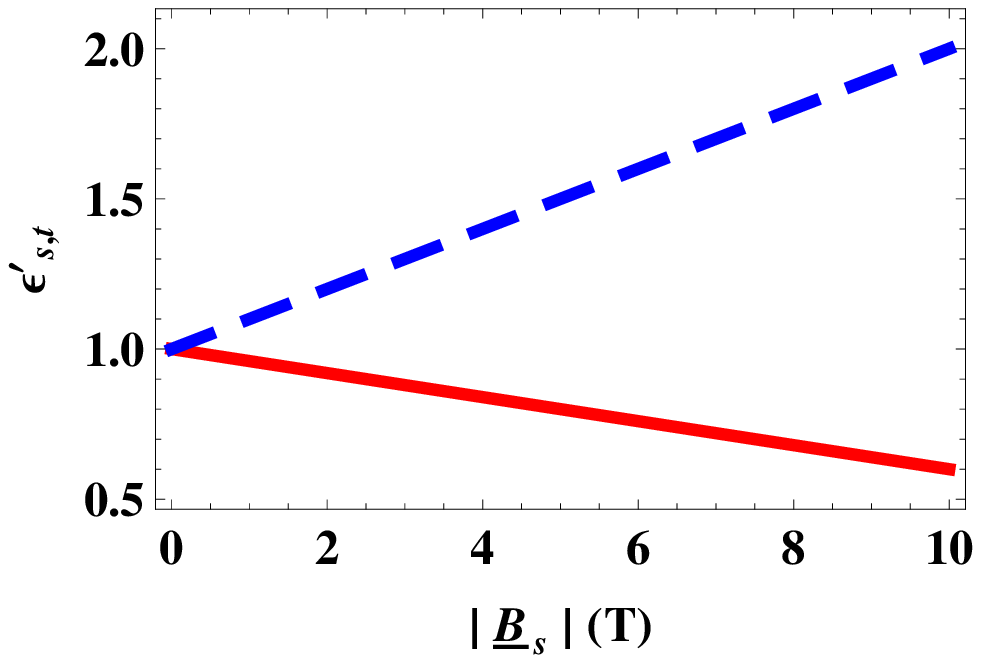} \caption{\label{fig1}
 The relative permittivity parameters $\eps'_s$ (blue, dashed) and $\eps'_t$ (red, solid) plotted versus $| \#B_{\,s} |$ (T).  }
\end{figure}

Since  $a$ is exceedingly small, the value of $a |\#B_s|^2$ is also
exceedingly small in comparison with unity for typical values of
$|\#B_s|$. For example, in the PVLAS experiment $|\#B_s| = 5$ T
typically \c{Zav}, which yields $a |\#B_s|^2 = 1.656 \times
10^{-24}$. Accordingly, the degree of anisotropy represented by the
constitutive dyadics \r{CDs} is also exceedingly small. In order to
achieve degrees of anisotropy that could be realistically attained
in a controlled manner for a practical simulation of QED vacuum, we
implement the  affine  transformation
\begin{equation}
\#x \mapsto \#x' \equiv \=J \. \#x \l{affine}
\end{equation}
of the spatial coordinates.
The transformation dyadic
\begin{equation}
\=J = p \, \le \=I- \hat{\#B}_s \hat{\#B}_s \ri + q \,\hat{\#B}_s
\hat{\#B}_s
\end{equation}
employs
\begin{equation}
\left.
\begin{array}{l}
\displaystyle{ p = \sqrt{\frac{ \le 1-8 a |\#B_s|^2 \ri  \le 1+20 a
|\#B_s|^2
\ri}{\le 1- 2 \delta |\#B_s| \ri \le1+ 5 \delta |\#B_s| \ri}} }\vspace{8pt}\\
\displaystyle{q = \frac{  1-8 a |\#B_s|^2 }{ 1 - 2 \delta |\#B_s|}}
\end{array}
\right\}
\end{equation}
and the scalar parameter $\delta > 0$.  For definiteness, we fix
$\delta = 0.02$.
 Thus, the
affine-transformed relative permittivity and permeability dyadics
are given as \c{PiO_cloaking}
\begin{eqnarray}
\=\eps' &\equiv& \frac{1}{\det \=\eps} \, \=J \. \=\eps  \. \=J^T \\
&=& \le 1- 2 \delta |\#B_s| \ri  \le \=I- \hat{\#B}_s \hat{\#B}_s
\ri + \le 1+ 5 \delta |\#B_s|\ri \hat{\#B}_s \hat{\#B}_s
\\ &\equiv& \eps'_t  \le \=I- \hat{\#B}_s \hat{\#B}_s \ri + \eps'_s \hat{\#B}_s \hat{\#B}_s
\end{eqnarray}
and
\begin{eqnarray}
\=\mu' &\equiv& \frac{1}{\det \=\mu} \, \=J \. \=\mu  \. \=J^T \\
&=& \frac{ 1- 2 \delta |\#B_s|}{ \le 1 - 8 a |\#B_s|^2\ri^2 }  \,
\le \=I- \hat{\#B}_s \hat{\#B}_s \ri + \frac{ 1+ 5 \delta |\#B_s|}{1
- 4 a |\#B_s|^2
\le 1 + 120 a |\#B_s|^2 \ri } \, \hat{\#B}_s \hat{\#B}_s \l{mu2} \\
&\equiv& \mu'_t  \le \=I- \hat{\#B}_s \hat{\#B}_s \ri + \mu'_s
\hat{\#B}_s \hat{\#B}_s .
\end{eqnarray}
Notice that for the range $|\#B_s| \in \le 0,  10 \ri$ T,  the
denominators of both terms on the right side of Eq.~\r{mu2} are both
approximately equal to unity, and therefore $\=\eps' \approx
\=\mu'$. The components $\eps'_s$ and $\eps'_t$ are linearly
dependent  upon $|\#B_s|$, as illustrated in Fig.~\ref{fig1}.  The
plots of $\mu'_{s,t}$ versus $|\#B_s|$ are practically identical to
those of $\eps'_{s,t}$.

\section{Simulation as a homogenized composite material}

Let us now turn to the question: How can one specify an HCM which is
a uniaxial dielectric-magnetic material with relative permittivity
dyadic $\=\eps_{\,HCM} \equiv \=\eps'$ and relative permeability
dyadic $\=\mu_{\,HCM} \equiv \=\mu'$? In order to answer this
question, we make use of the well-established Bruggeman
homogenization formalism \cite{WLM1997,EAB}.

Suppose we consider the homogenization of four component materials,
labelled $a$, $b$, $c$ and $d$. Two of the components ($a$ and $b$,
say) are isotropic dielectric materials while the other two ($c$ and
$d$)  are isotropic magnetic materials. Thus, the component
materials are specified by
\begin{itemize}
\item[(i)] the relative permittivities $\eps_a$, $\eps_b$,
$\eps_c$ and $\eps_d$, with $\eps_c = \eps_d = 1$; and
\item[(ii)] the relative
permeabilities $\mu_a$, $\mu_b$, $\mu_c$ and $\mu_d$, with $\mu_a =
\mu_b = 1$.
\end{itemize}
The four component materials are
randomly distributed with  volume fractions $f_a$, $f_b$, $f_c$,
$f_d \in \le 0,1 \ri$, with $f_d = 1 - f_a - f_b - f_c$. All four
component materials consist of  identically oriented spheroidal
particles. The symmetry axis for all these spheroidal particles
lies parallel to $\hat{\#B}_s$. Accordingly, the surface of each
spheroid relative to its centre is prescribed by the vector
\begin{equation}
\#r_{\,s} = \rho_\ell \, \=U_{\, \ell} \cdot \hat{\#r},
\end{equation}
wherein the shape dyadic
\begin{equation}
\=U_{\, \ell} =   \le \=I- \hat{\#B}_s \hat{\#B}_s \ri + U_\ell
\hat{\#B}_s \hat{\#B}_s, \qquad \qquad \le \ell = a, b, c, d \ri,
\end{equation}
is real symmetric \c{Lakh2000} and positive definite,
the radial unit vector is $\hat{\#r}$, and the linear measure
$\rho_\ell$ is required to be small compared to the electromagnetic
wavelengths under consideration. The shape parameter $U_\ell > 1$
for prolate spheroids and $0 < U_\ell < 1$ for oblate ones.

Under the Bruggeman homogenization formalism, the corresponding HCM
is a uniaxial dielectric-magnetic material, specified by  relative
permittivity
 and
permeability dyadics of the  form
\begin{equation}
\=\tau_{\,HCM} =  \tau^{HCM}_t  \le \=I- \hat{\#B}_s \hat{\#B}_s \ri
+ \tau^{HCM}_s \hat{\#B}_s \hat{\#B}_s  , \qquad \le \tau = \eps,
\mu \ri.
\end{equation}
For the particular  case of the uniaxial dielectric-magnetic HCM
involved here, full details of the numerical process of computing
the dyadics $\=\eps_{\,HCM}$ and $\=\mu_{\,HCM}$, from a knowledge
of $\eps_{a,b,c,d}$, $\mu_{a,b,c,d}$, $U_{a,b,c,d}$ and
$f_{a,b,c,d}$,
 are provided elsewhere \c{ML_PRB}.

Conventionally, homogenization formalisms are used to estimate the
   constitutive parameters of HCMs, based on  a
knowledge of the constitutive and morphological parameters of their
component materials and their volume fractions. In contrast,
  here our goal is to estimate the
constitutive and morphological parameters as well as the volume
fractions of the component materials which  give rise to a HCM such
that $\=\eps_{\,HCM}$ coincides with $\=\eps'$ and $\=\mu_{\,HCM}$
coincides with $\=\mu'$. We do so via an inverse implementation of
the
 Bruggeman formalism. Formal expressions
of the inverse Bruggeman formalism are available \c{WSW_MOTL}, but
in some instances these  can be ill-defined \c{Cherkaev}. In
practice, the inverse formalism may be more effectively implemented
by direct numerical methods  \c{ML_JNP}. Note that certain
constitutive parameter regimes
 have been identified as problematic for the inverse Bruggeman formalism \c{SSJ_TGM}, but these  regimes
are not the same as
 those
 considered here.

We consider the following three different implementations of the
inverse Bruggeman formalism. In each implementation,  four scalar
parameters are to be determined.
\begin{itemize}
\item[I.] The relative permittivities $\eps_{a,b}$ and the relative permeabilities
 $\mu_{c,d}$ are assumed to be known, and all spheroidal particles have the same shape, i.e.,  $U_a = U_b = U_c = U_d \equiv U$.
 We then determine the common shape parameter $U$ and the volume
 fractions $f_a$, $f_b$ and $f_c$.
 \item[II.] The relative permittivities $\eps_{a,b}$ and the relative permeabilities
 $\mu_{c,d}$ are assumed to be known, and  the volume fractions $f_{a,b,c}$ are
 fixed.
 We then determine the  shape parameters  $U_a$, $U_b$, $U_c$ and $U_d$.
\item[III.] The shape parameters $U_{a,b,c,d}$ and the volume fractions $f_{a,b,c}$ are
fixed.
 We then determine  the relative permittivities $\eps_{a,b}$ and relative permeabilities
 $\mu_{c,d}$.
\end{itemize}

To describe the inversion of the Bruggeman formalism, let us focus
on implementation  I as a representative example, the inversion
processes for implementations II and III being analogous. Suppose
that  $\lec \breve{\eps}^{HCM}_{s}, \breve{\eps}^{HCM}_t,
\breve{\mu}^{HCM}_s, \breve{\mu}^{HCM}_t \ric $ are  forward
Bruggeman estimates of the HCM's relative permittivity and relative
permeability parameters which are computed  for physically
reasonable ranges of the parameters $U$ and $f_{a,b,c}$; i.e.,
  $ U \in
\le U^-, U^+ \ri$ and  $f_{a,b,c} \in \le f^-_{a,b,c}, f^+_{a,b,c}
\ri$. Next:
\begin{itemize}
\item[(1)] Let $ f_a = \le f^-_a + f^+_a \ri /2$, $ f_b = \le f^-_b + f^+_b \ri /2$,
and $ f_c = \le f^-_c + f^+_c \ri /2$. For all  $U \in \le U^-, U^+
\ri$, determine the value $U^\dagger$  which yields the minimum
value of the scalar quantity
 \begin{eqnarray} \Delta &=& \Big[\le
\frac{\breve{\eps}^{HCM}_s  - \eps'_s}{\eps'_s} \ri^2 + \le
\frac{\breve{\eps}^{HCM}_t - \eps'_t}{\eps'_t} \ri^2 + \le
\frac{\breve{\mu}^{HCM}_s - \mu'_s}{\mu'_s} \ri^2 + \le
\frac{\breve{\mu}^{Br}_t - \mu'_t}{\mu'_t} \ri^2 \Big]^{1/2}.
\end{eqnarray}
\item[(2)] Let $U = U^\dagger$,
 $ f_b = \le f^-_b + f^+_b \ri /2$,
and $ f_c = \le f^-_c + f^+_c \ri /2$.
 For all  $f_a \in \le f_a^-, f_a^+ \ri$, determine the
value $f_a^\dagger$  which yields the minimum value of $\Delta$.
\item[(3)] Let $U = U^\dagger$, $ f_a = f_a^\dagger$, and $ f_c = \le f^-_c + f^+_c \ri
/2$. For all  $f_b \in \le f_b^-, f_b^+ \ri$, determine the value
$f_b^\dagger$  which yields the minimum value of  $\Delta$.
\item[(4)] Let $U = U^\dagger$, $ f_a = f_a^\dagger$, and $ f_b = f_b^\dagger$.
 For all  $f_c \in \le f_c^-, f_c^+ \ri$, determine the value
$f_c^\dagger$  which yields the minimum value of   $\Delta$.
\end{itemize}
The steps (1)--(4) are then repeated, with $f_a^\dagger$,
$f_b^\dagger$, and $f_c^\dagger$ being the fixed values of
$f_{a,b,c}$
 in step (1),   $f_b^\dagger$ and  $f_c^\dagger$ being the fixed values of
$f_{b,c}$ in step (2), and  $f_c^\dagger$  being the fixed value of
$f_{c}$ in step (3),
 until the value of $\Delta$ becomes acceptably
small.

\section{Numerical illustrations}

Numerical illustrations of the implementations I--III are provided
in Figs.~\ref{fig2}--\ref{fig4}. For all results presented, the
degree of convergence of the numerical schemes which provide the
inverse Bruggeman estimates
 was $< 1 \%$, and in most instances this value was $< 0.1 \%$.

\begin{itemize}
\item[I.] For Fig.~\ref{fig2}, the constitutive parameters of the component
materials were taken to be
 $\epsilon_a = 4$, $\epsilon_b = 0.3$, $\mu_c = 3.4$ and $\mu_d = 0.4$.
The computed common shape parameter $U$ and volume fractions
$f_{a,b,c}$ are plotted versus $| \#B_s |$. While the volume
fractions vary little as $| \#B_s |$ is increased from 1 to 2.5 T,
the common shape parameter increases exponentially.

\item[II.] The constitutive parameters of the component materials were again
taken to be
 $\epsilon_a = 4$, $\epsilon_b = 0.3$, $\mu_c = 3.4$ and $\mu_d =
 0.4$ for Fig.~\ref{fig3}. In addition, the volume fractions were
 fixed at $f_a = 0.15$, $f_b = 0.25$ and $f_c = 0.21$. The computed four
 shape parameters $U_{a,b,c,d}$ are plotted versus $| \#B_s |$. All
 four shape parameters increase uniformly as $| \#B_s |$ is increased from 1 to 2
 T. This reflects the fact that the degree of anisotropy  of the HCM is
 required to increase as $| \#B_s |$ increases.

\item[III.] Lastly, in Fig.~\ref{fig4} the common shape parameter is fixed at
 $U = 5$ while the volume fractions are fixed at $f_{a,b,c} = 2.5$.
The computed constitutive parameters $\eps_{a,b}$ and $\mu_{c,d}$
are plotted versus $| \#B_s |$. In this case, $\eps_a$ turns out to
be approximately the same as $\mu_c$. And similarly $\eps_b$ turns
out to be approximately the same as $\mu_d$. While $\eps_a$ and
$\mu_c$ decrease uniformly as $| \#B_s |$ is increased from 1 to 3
 T, the opposite is true of $\eps_b$ and $\mu_d$.
 \end{itemize}

\section{Closing remarks}

By means of the inverse Bruggeman formalism, an HCM may be specified
which is electromagnetically equivalent to the QED vacuum subject to
a
 spatial  affine transformation. The affinely transformed
QED vacuum retains the same uniaxial dielectric-magnetic form  as
the un-transformed QED vacuum, but the degee of anisotropy is
greatly exaggerated by means of the affine transformation. By
reversing the transformation represented by eq.~\r{affine}, the
properties of QED vacuum may be inferred from those of the HCM.

For illustration, the inverse homogenization formulation presented
here was based on four isotropic component materials. However, the
desired HCM could also be realized by alternative inverse
homogenization formulations. For example, the HCM could arise from
only two component materials. These two components materials could
be either both isotropic dielectric-magnetic materials distributed
as oriented spheroidal particles or both uniaxial
dielectric-magnetic materials (with parallel symmetry axes)
distributed as spherical particles \c{EAB,MW_JOPA}. However, the
four-component formulation presented here involves the simplest of
component materials and allows a large degree of freedom in choosing
their constitutive parameters.

\begin{figure}[!ht]
\centering
\includegraphics[width=4.3in]{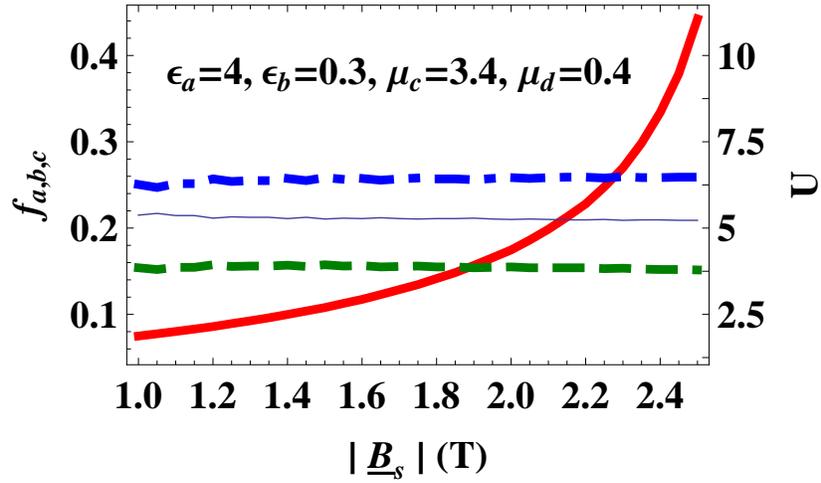}
 \caption{\label{fig2} Implementation I.
 The common shape parameter $U$ (thick solid,  red)  and volume fractions $f_{a}$
  (dashed, green), $f_b$ (broken dashed, blue), and $f_c$ (thin solid, blue)
  plotted versus $ |\#B_{\,s}|$ (T).  The relative permittivities
 $\epsilon_a = 4$, $\epsilon_b = 0.3$, $\mu_c = 3.4$ and $\mu_d = 0.4$.  }
\end{figure}

\begin{figure}[!ht]
\centering
\includegraphics[width=4.3in]{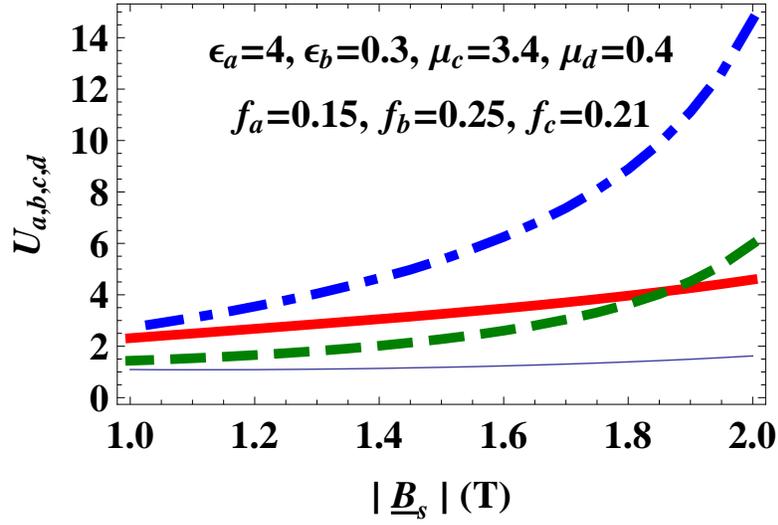}\\
 \caption{\label{fig3} Implementation II.
  The shape parameters $U_{a}$ (thick solid,  red), $U_{b}$ (dashed, green), $U_{c}$ (broken dashed, blue), and $U_{d}$  (thin solid, blue)
    plotted versus $ |\#B_{\,s}|$ (T). The relative permittivities
 $\epsilon_a = 4$, $\epsilon_b = 0.3$, $\mu_c = 3.4$ and  $\mu_d = 0.4$; and the volume fractions $f_{a} = 0.15$, $f_{b} = 0.25$ and $f_{c} = 0.21$.
   }
\end{figure}

\begin{figure}[!ht]
\centering
\includegraphics[width=4.3in]{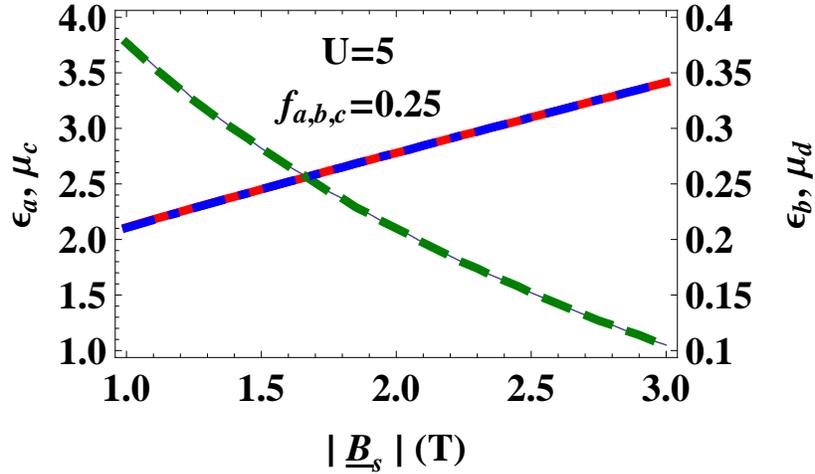}
 \caption{\label{fig4} Implementation III.
The  relative permittivities $\epsilon_{a}  $ (thick solid, red) and
$\epsilon_{b}$ (dashed, green)
 and the relative permeabilities $\mu_{c} $ (broken dashed, blue) and $\mu_{d}$ (thin solid, blue)
 plotted versus $ |\#B_{\,s}|$ (T).
The  shape parameter $U=U_{a,b,c,d} = 5$ and volume fractions
$f_{a,b,c} = 0.25$. }
\end{figure}

Finally, let us note that the relative permittivities and relative
permeabilities of the component materials needed for the HCM, as
presented in Figs.~\ref{fig2}--\ref{fig4}, are not at all
infeasible. Indeed, present-day technology allows for the
possibility of
 materials with a considerably wider range of
 constitutive parameters
 to be engineered
 \c{Alu,Lovat,Cia_PRB}.

\vspace{5mm}

\noindent {\bf Acknowledgment:}  AL thanks the Charles Godfrey
Binder Endowment at Penn State for partial financial support of his
research activities. 

\newpage

\newpage

\end{document}